\documentclass[sigconf, nonacm]{acmart}

\usepackage{sparklines}
\usepackage[export]{adjustbox}
\usepackage{subfig}
\usepackage{xspace}
\usepackage{listings}
\lstdefinestyle{mystyle}{
    basicstyle=\ttfamily,
    breakatwhitespace=false,
    breaklines=true,
    captionpos=b,
    keepspaces=true,
    numbers=left,
    numbersep=5pt,
    showspaces=false,
    showstringspaces=false,
    showtabs=true,
    tabsize=2
}
\usepackage{courier}
\usepackage{bbding}
\usepackage{etoolbox}
\usepackage{calc}
\makeatletter
\pretocmd{\@chooseSymbol}{\raisebox{-.5ex}[\height-.5ex][0pt]}{}{}
\makeatother
\usepackage{array}
\newcolumntype{P}[1]{>{\centering\arraybackslash}m{#1}}

\usepackage{xcolor}
\usepackage{soul}
\usepackage{marginnote}
\usepackage{mdframed}
\usepackage{multirow}
\usepackage{amsmath}
\usepackage{enumitem}

\newcommand{\tinyskip}{\vspace{3pt}}
\newcommand{\mypar}[1]{\tinyskip\noindent\textbf{#1.}\xspace}

\newcommand{\F}{\mbox{Fig.\hspace{0.25em}}}

\newenvironment{myitemize}{%
\begin{itemize}[leftmargin=1em, itemsep=.1em, parsep=.1em, topsep=.1em,
    partopsep=.1em]}
{\end{itemize}}

\newenvironment{myenumerate}{%
\begin{enumerate}[leftmargin=1em, itemsep=.1em, parsep=.1em, topsep=.1em,
    partopsep=.1em]}
{\end{enumerate}}

\newenvironment{structure*}{\color{blue}\begin{myenumerate}}{\end{myenumerate}}

\theoremstyle{definition}
\newtheorem{definition}{Definition}

\theoremstyle{theorem}
\newtheorem{exmp}{Example}

\sethlcolor{yellow}

\newcommand\vldbdoi{XX.XX/XXX.XX}
\newcommand\vldbpages{XXX-XXX}
\newcommand\vldbvolume{14}
\newcommand\vldbissue{1}
\newcommand\vldbyear{2020}
\newcommand\vldbauthors{Siyuan Xia, Chris Zhu, Tapan Srivastava, Bridget Fahey, Raul Castro Fernandez}
\newcommand\vldbtitle{\shorttitle} 

\newcommand\vldbpagestyle{plain}

\begin{document}

\title{Programmable Data Sharing}
\title{Programmable Dataflows: A \emph{Contract} Abstraction and API to Enable Data Sharing on Data Escrows}
\title{Programmable Dataflows: A Data Management Abstraction and Programming Model to Enable Data Sharing}
\title{Programmable Dataflows: Abstraction and Programming Model for Data Sharing}

\author{Siyuan Xia, Chris Zhu, Tapan Srivastava, Bridget Fahey*, Raul Castro Fernandez}
\affiliation{\vspace{0.1cm}
\institution{The University of Chicago, The University of Chicago Law School*}
}
\email{{stevenxia, chz, tapansriv, bridget.fahey, raulcf}@uchicago.edu}

\begin{abstract}


Data sharing is central to a wide variety of applications such as fraud detection, ad matching, and research. The lack of data sharing abstractions makes the solution to each data sharing problem bespoke and cost-intensive, hampering value generation. In this paper, we first introduce a data sharing model to represent every data sharing problem with a sequence of dataflows. From the model, we distill an abstraction, the \textit{contract}, which agents use to communicate the intent of a dataflow and evaluate its consequences, before the dataflow takes place. This helps agents move towards a common sharing goal without violating any regulatory and privacy constraints. Then, we design and implement the contract programming model (CPM), which allows agents to program data sharing applications catered to each problem's needs.

Contracts permit data sharing, but their interactive nature may introduce inefficiencies. To mitigate those inefficiencies, we extend the CPM so that it can save intermediate outputs of dataflows, and skip computation if a dataflow tries to access data that it does not have access to. In our evaluation, we show that 1) the contract abstraction is general enough to represent a wide range of sharing problems, 2) we can write programs for complex data sharing problems and exhibit qualitative improvements over other alternate technologies, and 3) quantitatively, our optimizations make sharing programs written with the CPM efficient.



\end{abstract}

\maketitle
\let\thepage\relax 
\pagestyle{plain}

\pagestyle{\vldbpagestyle}
\begingroup\small\noindent\raggedright\textbf{PVLDB Reference Format:}\\
\vldbauthors. \vldbtitle. PVLDB, \vldbvolume(\vldbissue): \vldbpages, \vldbyear.\\
\href{https://doi.org/\vldbdoi}{doi:\vldbdoi}
\endgroup
\begingroup
\renewcommand\thefootnote{}\footnote{\noindent
This work is licensed under the Creative Commons BY-NC-ND 4.0 International License. Visit \url{https://creativecommons.org/licenses/by-nc-nd/4.0/} to view a copy of this license. For any use beyond those covered by this license, obtain permission by emailing \href{mailto:info@vldb.org}{info@vldb.org}. Copyright is held by the owner/author(s). Publication rights licensed to the VLDB Endowment. \\
\raggedright Proceedings of the VLDB Endowment, Vol. \vldbvolume, No. \vldbissue\ %
ISSN 2150-8097. \\
\href{https://doi.org/\vldbdoi}{doi:\vldbdoi} \\
}\addtocounter{footnote}{-1}\endgroup

\section{Introduction}
\label{sec:introduction}


Data sharing is a central process of our socioeconomic environment. Examples of how data sharing impacts the economy include the multi-billion ad-tech business~\cite{iab_report} and the data workers who share data to create reports, perform better decision-making, and complete other value-generating tasks~\cite{jasonreport}. Examples beyond those with direct economic impact include health organizations sharing data to improve patient care~\cite{hie2018realizing,jha2008towards,rogers2022vaultdb,geva2023collaborative,froelicher2021truly}, financial institutions sharing data to improve fraud detection~\cite{tnml, austrac,acfe_report}, and citizens sharing data with their governments to facilitate resource allocation~\cite{australiasharingagreement}. Today, we think of the above examples as different problems that require bespoke solutions that are compliant with regulations and privacy constraints. In summary, we lack technical infrastructure~\cite{bidenorder, nsfpdasp} for data sharing, making data sharing solutions cost-intensive, thus hampering value generation.



In this paper, we identify a lack of \emph{abstractions} as the culprit behind cost-inefficient data sharing solutions. It is difficult to unearth what is common across data sharing problems and too easy to get mired in details of compliance, regulation, and privacy, which are hyper-specific to the local regulations and the organizations' bylaws. Against this backdrop, our paper contributes:


\begin{myitemize}

\item A new \emph{data sharing model} that represents each data sharing problem as a sequence of \emph{dataflows}---data exchanges between participants. While some dataflows help agents achieve their sharing goal and are thus desirable, some dataflows violate agents' constraints (e.g. leaking sensitive data) and should be avoided.

\item A main challenge of data sharing is that agents lack information to assess whether a dataflow is desirable before it takes place. To preclude adverse consequences, agents default to not sharing. We introduce a new \emph{contract abstraction} that bounds such consequences by making it explicit \textit{who} contributes data, \textit{what computation} takes place on that data, \textit{who} receives the result, and under \textit{what conditions}. Importantly, it provides this information before an intended dataflow takes place, thus addressing the challenge by helping agents make an informed decision on whether to allow the dataflow.

\item We design and implement a contract programming model (CPM) that developers use to write data sharing applications. Using the CPM vastly reduces the cost of addressing data sharing problems, when access to software is available.

\end{myitemize}

Together, the \emph{contract} abstraction and programming model enable \textsc{programmable dataflows}, a paradigm by which agents can analyze a sharing problem, identify what dataflows should (or should not) take place, and write a program that permits participating agents to share data in a controlled way that respects the desired dataflows, while complying with regulations, compliance rules, and privacy preferences. Consider the following illustrative example:

\begin{exmp}[\bf{Financial Fraud Detection}]
\label{ex:example}
Banks are interested in pooling credit card transaction data to train more accurate fraud detection models~\cite{awoyemi2017credit, acfe_report}, but they are generally not allowed to reveal their plaintext data. Moreover, it is not commercially interesting to pool their data, unless every bank has the guarantee that the joint model benefits themselves, instead of only helping others. Thus, the banks may participate in data sharing if the following requirements are met: i) no raw data will ever be disclosed directly; ii) each bank contributes sufficient training data; and iii) the joint model can achieve some minimum accuracy on each banks' test data. 



\end{exmp}

Today, addressing this data sharing problem requires concerted efforts, such as those by the TNML~\cite{tnml} and Austrac~\cite{austrac} consortia. Unfortunately, these are one-off, hard to replicate, and even harder to transfer to other data sharing problems. 

In contrast, \textsc{programmable dataflows} permit a developer to implement a program using the CPM, served via a data escrow, that lets participants share data while controlling the released dataflows. Participants \emph{interact} with this program by proposing \emph{contracts} that express their intent, revising contracts when they are incompatible with privacy preferences or compliance rules, and releasing dataflows only when contracts are approved. The interactions stemming from contract negotiations are useful when preparing data (e.g., align schemas, standardize formats, etc.), so that participants can agree on a set format.

\textsc{Programmable dataflows} also introduce new challenges. The contract abstraction \emph{breaks} a program's normal computation flow, because agents need to be involved to manage contracts. This prevents common data management optimizations needed for performance, such as saving and skipping computation. To that end:

\begin{myitemize}

\item We augment the programming model to permit reusing the intermediate output of a dataflow without making it available to anyone---which would otherwise break the contract. We implement short-circuiting to skip computation whenever a dataflow is about to access data it does not have access to. Together, these features mitigate those performance inefficiencies.

\item An unavoidable consequence of representing data sharing problems programmatically is that it is necessary to suspend computation whenever an agent must intervene to decide whether to release data. We implement mechanisms to reduce human interventions.

\end{myitemize}

In the evaluation, we ask three questions, which represent the main hypotheses of our work. First, we explore, qualitatively, whether the contract abstraction is general enough to represent varied and common data sharing problems. Second, we investigate whether we can write data sharing programs for 3 instances of complex data sharing problems (including the fraud detection problem above, a national health consortium for medical data sharing \cite{schmidt2015danish}, and ad matching done usually with a data clean room \cite{iab_report}). We qualitatively compare the implementation to other alternative technologies for data sharing. Third, we quantitatively evaluate the performance of data sharing programs written using the CPM.

To the best of our knowledge, this is the first paper that concentrates, holistically, on the challenge of data sharing. Most existing works concentrate on the regulatory problem (the legal literature \cite{walsh2022multi, helminger2022multi,reyes2018won,edemekong2018health}), the incentive problem (the economics literature \cite{oecd_enhancing_2019, agarwal2019marketplace, castro2023data}, the preparation problem (the data discovery and integration literature \cite{ranbaduge2022privacy,clifton2004privacy,galhotra2023metam,lenzerini2002data,de2010practical, chen2017fast, hazay2017scalable, kolesnikov2017practical}) or the execution problem \cite{nguyen2023blockchain, bater2017smcql, bater2020saqe, zheng2021cerebro, volgushev2019conclave,hunt2018ryoan}. Most works on the execution problem assume that agents are already willing to share, and they focus on securely executing data exchange. Our work differs from point solutions in that it tackles the ``lack of abstractions'' problem, by making a large number of sharing problems programmable with the contract abstraction and a programming model. Overall, our paper contributes to ongoing calls (from government~\cite{bidenorder} and funding agencies~\cite{nsfpdasp}) for improving technical data sharing infrastructures.

The rest of the paper is structured as follows. Section~\ref{sec:model} introduces a novel data sharing model based on dataflows, and identifies an approach to data sharing. Section~\ref{sec:contract} presents and illustrates the contract abstraction. Section~\ref{sec:api} introduces a programming model to the contract abstraction. Section~\ref{sec:evaluation} presents the evaluation results and Section~\ref{sec:conclusions} the conclusions.

\section{The Data Sharing Model}
\label{sec:model}

In this section, we present a data sharing model we use to systematically think of data sharing problems.

\mypar{Overview of the Model} In the model, data sharing is represented as a series of \emph{data sharing states}. A state describes which \emph{agent} has access to what \emph{data}. Agents have a preference to achieve certain states (\emph{goal states}) but also need to prevent certain states (\emph{constraint states}), for example because of regulatory, compliance, and privacy reasons. Agents move from one state to another via a \emph{dataflow}. 


\subsection{Concepts of the Data Sharing Model}



\begin{definition}[\bf{Agent}]
\label{def:agent}
\emph{An agent, $a_i$, is an entity with the agency to participate in data sharing.}
\end{definition}

In the example, agents are the banks. The concept of an \textit{agent} encompasses various entities in data sharing literature, including data owners \cite{syft_2023}, data providers \cite{nguyen2023blockchain, databricks_2022}, data consumers \cite{databricks_2022}, data scientists \cite{syft_2023}, and data requesters \cite{nguyen2023blockchain}.

\begin{definition}[\bf{Data Element}]
\label{def:dataelement}
\emph{A data element $d_i$ represents some logical grouping of data.} 
\end{definition}

The granularity of data elements is specific to each sharing problem; it may be a text document, image, database, table's cell, etc. Data elements may be derived from other data elements via a function $f()$, e.g., a classifier ML model is a data element built by applying a training function over some input training data (the input data element). 



\begin{figure}%
    \centering
    \includegraphics[width=0.82\columnwidth]{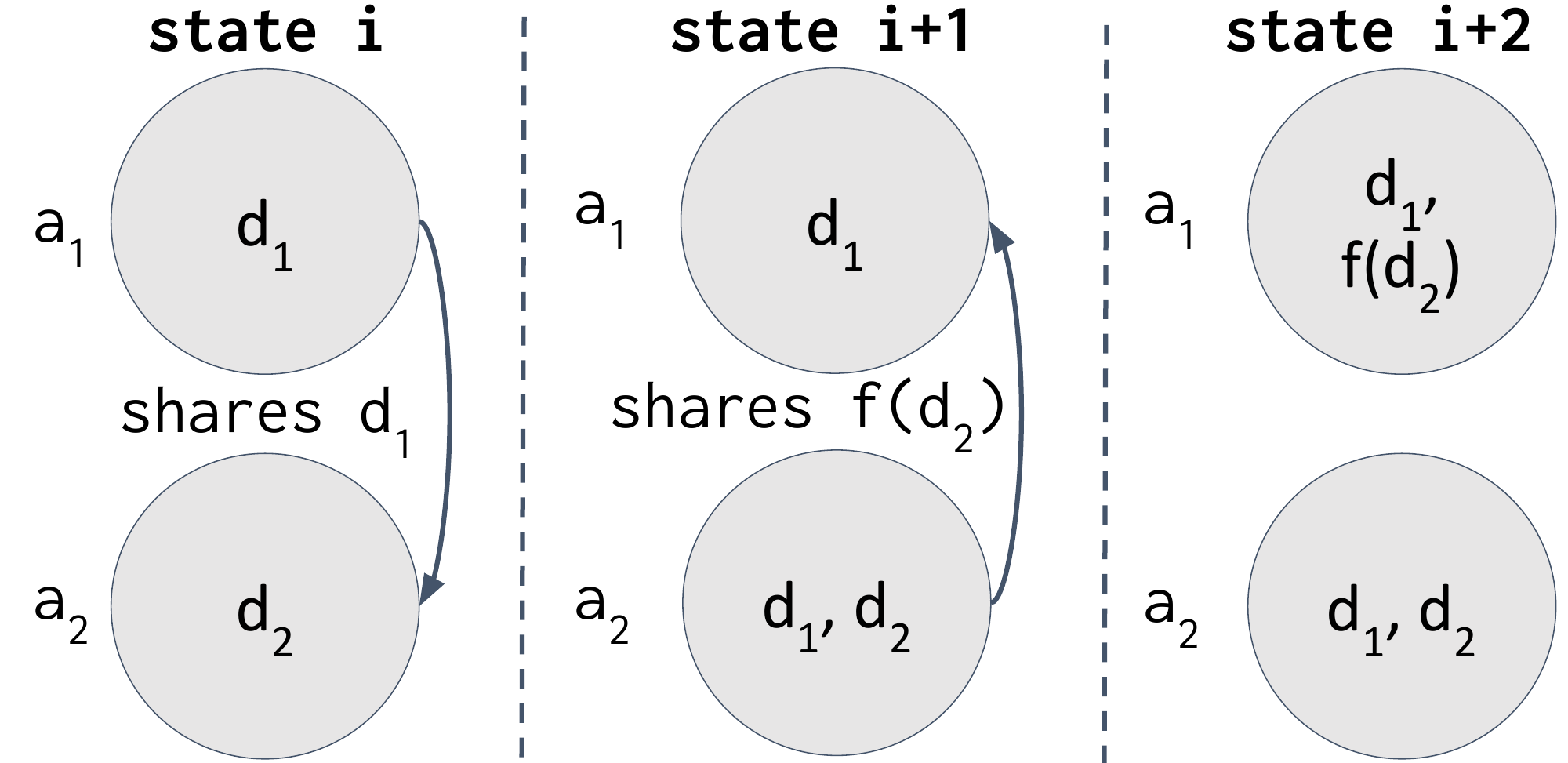}
    \caption{Illustration of three consecutive data sharing states.}%
    \label{fig:state}%
\end{figure}

\begin{definition}[\bf{Data Sharing State}]
\label{def:state}

\emph{A data sharing state, $q$, is defined by a set of agents $A$, a set of data elements $D$, and a correspondence $s \colon A \to D$, that yields the subset of $D$ that is accessible to each agent. That is, $q = \langle A, D, s \rangle$.}

\end{definition}

Initially, there exists a set of agents $A_0$ with a set of data elements $D_0$. We denote this \textit{initial state} by $q_0$. Whenever a data element is shared with another agent, there is a \textit{transition} to a new state; the details of the transition are explained in the following subsection. Data sharing is a series of transitions from states $q_0, q_1, \cdots, q_i, \cdots$. For state $i$, $q_i = \langle A_i, D_i, s_i \rangle$. Figure \ref{fig:state} shows three sharing states. In state i, $A_i$ = \{$a_1$, $a_2$\}, $D_i$ = \{$d_1$, $d_2$\}, $s_i(a_1)$ = \{$d_1$\}, and $s_i(a_2)$ = \{$d_2$\}. In state i+1, $A$ and $D$ remain unchanged, and $s_{\text{i+1}}(a_2)$ = \{$d_1, d_2$\}.

Transitions (i.e., data sharing) are explicitly triggered by agents. Agents want to reach some final state where they have access to certain data elements. In the banking example, each bank wants to access the joint model that achieves some minimum accuracy. Such desired final states are examples of \textit{goal states}.

\begin{definition}[\bf{Goal States}]
\label{def:gs}

\emph{For an agent $a_i\in A_i$, their goal states are the set of states $q^g_i$ where $a_i$ gets access to the data elements that they want to access.}

\end{definition}

Each agent has their own set of goal states. For a set of agents, their \textit{common goal states} $q^G$ are the intersection of each individual's goal states: $q^G = \{q : q \in q^g_1 \cap q^g_2 \cap \cdots \cap q^g_m\}$. Reaching a common goal state is the desired outcome of a given data sharing problem. 


While agents would like to get access to data elements by entering their goal state, they would also like to avoid states where other agents have access to certain private, restricted data elements. As such, they want to avoid states that violate such restrictions. Such states are represented by \textit{constraint states}.

\begin{definition}[\bf{Constraint State}]
\label{def:cs}

\emph{
An agent $a_i$'s constraint states are the set of states $q^c_i$ that are disallowed by $a_i$. From $a_i$'s perspective, data sharing should never never reach $q^c_i$.
}

\end{definition}

Analogous to common goal states, the \textit{common constraint states} for a set of agents $q^C$ are the union of each individual's constraint states: $q^C = \{q : q \in q^c_1 \cup q^c_2 \cup \cdots \cup q^c_m\}$. 
In the fraud detection example, if a joint model is shared with any of the banks without meeting the accuracy requirement, then these banks have entered a common constraint state.



\subsection{Driving the Data Sharing Process}

Agents want to transition from an initial sharing state into a goal state while avoiding falling into a constraint state. Transitions between data sharing states are triggered by \emph{dataflows}, which indicate exchange of data elements among agents.



\begin{definition}[\bf{Dataflow}]

\emph{A dataflow $t_i()$ is a function that takes: a state $q_i$, a set of destination agents $A_{\text{dest}} \subseteq A_i$, a set of source data elements $D_{\text{src}} \subseteq D_i$, and a function $f$ to be applied on the source data elements. It produces the next state $q_{i+1}$, in which each destination agent $a_j \in A_{\text{dest}}$ gets access to a new data element $d^j$ $=$ $f(D_{\text{src}})$.}
    

    \[t_i(q_i, A_{\text{dest}}, D_{\text{src}}, f) = q_{i+1} = \langle A_i, D_{i+1}, s_{i+1} \rangle, \ \text{\emph{where}}\]
    \[
    D_{i+1} = D_i \cup d^j, \ \ \ 
    s_{i+1}(a_j) = 
    \begin{cases} 
        s_i(a_j) \cup d^j & a_j \in A_{\text{dest}} \\
        s_i(a_j) & \text{\emph{otherwise}}
    \end{cases}\]

\end{definition}

\noindent Dataflow is the central concept in the data sharing model. In Figure \ref{fig:state}, there are two dataflows. The first one gives $a_2$ access to $d_1$, and the second one gives $a_1$ access to $f(d_2)$. Each dataflow leads to a change in sharing state.

\begin{definition}[\bf{Data Sharing Goal}]
\emph{Agents $a_i \in A_i$ want to find a sequence of dataflows $t_0, t_1, ..., t_n$, which take them from the initial state $q_0$ to a goal state $q_i \in q^G$ (i.e., the agents have reached a \textit{common goal state}), such that none of the intermediate states is in one of the \textit{common constraint states} $q^C$ .}
\end{definition}

Unfortunately, at the beginning, agents often lack information about other agents' data elements, goal states, and constraint states. In the fraud detection example, the banks do not know if their data is compatible for the joint model to be trained. Also, they do not know how much information the other banks are willing to reveal for their data to be integrated. This lack of information makes it hard for agents to know if there exists a sequence of dataflows leading to their sharing goal, and what that sequence is.

Instead, agents learn more information as they make progress through the sharing process, considering one dataflow at a time. In general, an agent cannot unilaterally trigger a dataflow from one state to another without agreement from the other agents involved, because a dataflow that helps one agent move towards their goals may enter another's constraint state. To reach an agreement, agents must communicate their \emph{intent} of the dataflow and then assess whether the \emph{outcome} of the dataflow moves the sharing state closer to a goal state while avoiding their constraint states. 


\mypar{Main Challenge} 
What makes this challenging for agents is the resulting catch-22 problem: to approve a dataflow, agents want to see the outcome first, but to see the outcome they need to approve the dataflow. In the fraud detection example, the banks need the joint model to check if it meets each bank's accuracy requirement.

Agents do not want to commit to a dataflow without knowing whether the outcome takes them into a constraint state. This uncertainty results in dataflows that do not materialize, even in situations where they would advance the sharing state in a positive direction.

\section{The Contract Abstraction}
\label{sec:contract}

To address the above challenge, we introduce a new \textsf{contract} abstraction that lets agents \emph{understand the consequences of a dataflow \textbf{before} the dataflow materializes}, so they can decide whether to enact the dataflow. Then, we introduce operations using the contract that help agents work towards the sharing goal.

\subsection{Introducing the Contract Abstraction}


A contract determines: i) \textbf{what} data is shared; ii) \textbf{who} decides to share data with \textbf{whom}; and iii) \textbf{when} can data be shared. 

\mypar{What data is shared} Data shared in a contract is $f(DE)$, where $DE \subseteq D$ is some subset of the data elements, and $f$ is a transformation to be applied on $DE$ before it is shared. Examples of $f$ include training an ML model, running a SQL query over the data, and the identity function, in which case the original data is shared directly.

\mypar{Who decides to share with whom} A dataflow takes place if it is approved by a set of source agents, $A_{\text{src}}$. If it is approved, the outcome of the dataflow is that a set of destination agents, $A_{\text{dest}}$, gains access to $f(DE)$. Typically, $A_{\text{src}}$ needs to contain agents who contribute data elements to the dataflow, which is the set of all $a_i$, such that $s(a_i) \cap DE \neq \emptyset$. In addition to the agents who contribute data elements, $A_{\text{src}}$ may also contain other agents, such as compliance officers, who must sanction the proposed subsequent state. For example, a compliance officer may require that a dataflow can only happen if it does not leak any PII.

\mypar{When can data be shared} In every sharing problem, source agents need to consider common factors when they decide whether to share: what data is shared ($f(DE)$), and who receives the data ($A_{\text{dest}}$). In addition to these, $A_{\text{src}}$ may have additional requirements on what conditions need to be met for the data to be shared. We categorize those into \textsc{preconditions} and \textsc{postconditions}: 

\begin{myitemize}
    \item \textsc{precondition}: these conditions can be checked before $f(DE)$ is computed. In the fraud detection example, banks' precondition is to contribute sufficient data.
    \item \textsc{postcondition}: these conditions can only be checked after $f(DE)$ is computed. The banks' postcondition is that the joint ML model achieves a target accuracy on every bank's test data. This ensures that all banks benefit from sharing.
\end{myitemize}

We denote each source agent's decision on whether the data can be shared with a function, $M_{k} (A_{\text{dest}}, DE, f, C_{\text{pre}}, C_{\text{post}}) \rightarrow \{0,1\} \quad \\\forall a_k \in A_{\text{src}}$, with $1$ representing approval and $0$ denial. Then, a dataflow, $t_i(q_i, A_{\text{dest}}, DE, f)$, takes place if $M_k()=1, \forall a_k \in A_{\text{src}}$.



\begin{definition}[\bf{Contract}]

\emph{A contract is a 6-tuple:}
    \[\langle A_{\text{dest}}, DE, f, A_{\text{src}}, C_{\text{pre}},  C_{\text{post}} \rangle\]

    \noindent \emph{with a function $M_{k}(A_{\text{dest}}, DE, f, C_{\text{pre}}, C_{\text{post}}) \rightarrow \{0,1\}$ for every $a_k \in A_{\text{src}}$. $A_{\text{dest}}$ represents a set of destination agents, $DE$ represents a set of data elements as inputs to $f$,  $C_{\text{pre}}$ and $C_{\text{post}}$ represent the preconditions and postconditions, and $f(DE)$ is the data to be shared with all of the destination agents. $A_{\text{src}}$ is the set of agents who can approve the contract with $M_k$. The content of a contract is visible to all of $A_{\text{src}}$.}
\end{definition}


\subsection{Using the Contract Abstraction}

\begin{figure}%
    \centering
    \includegraphics[width=\columnwidth]{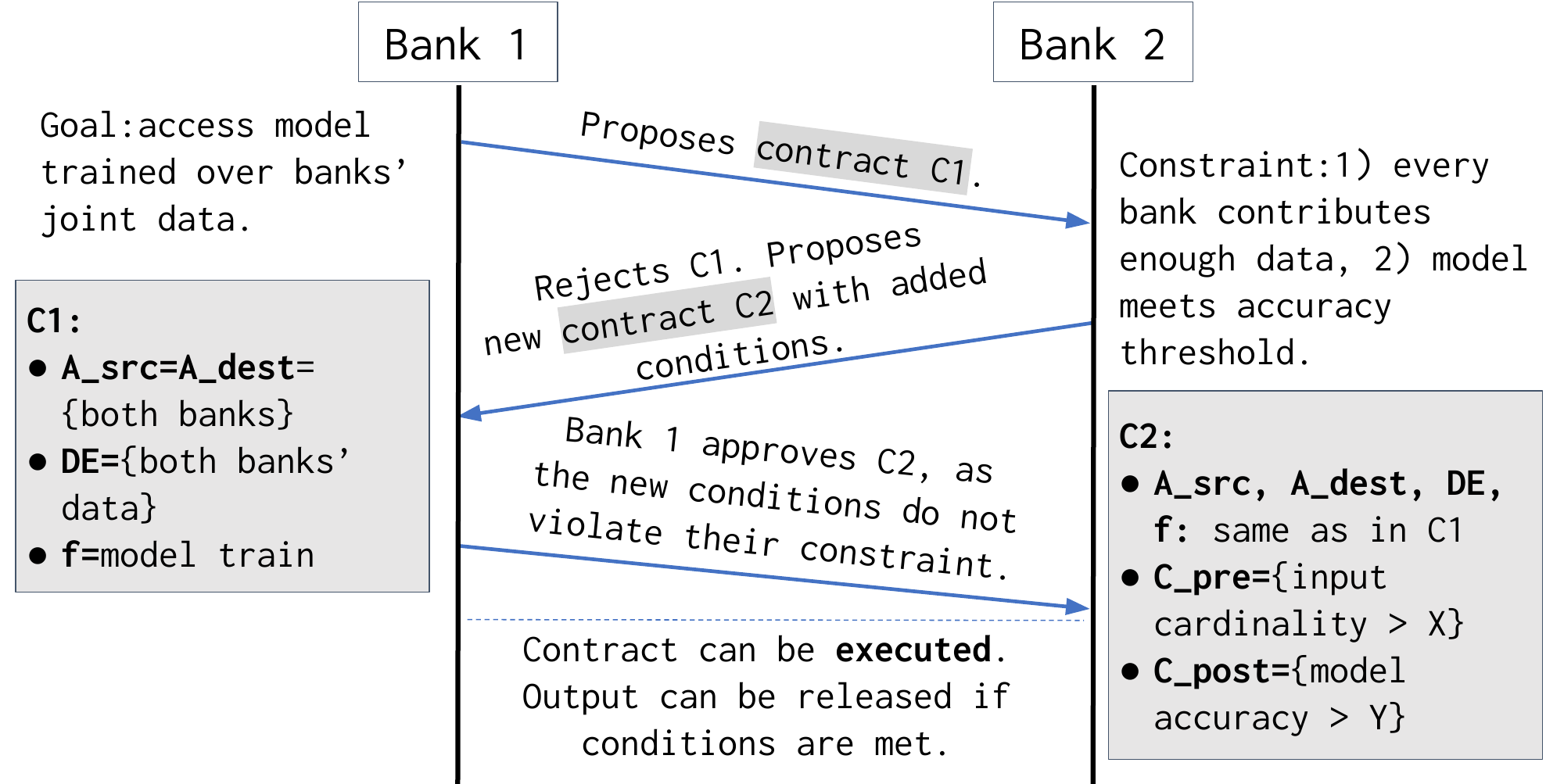}
    \caption{Examples of how contracts enable data sharing in the fraud detection example.}%
    \label{fig:contract}%
\end{figure}

The contract abstraction captures both the intent and the outcome of a dataflow, so agents can use contracts to solve their data sharing problems. They start with \textbf{proposing} a contract to the other agents. Once the contract is seen by the other agents, it may get \textbf{approved}. If a contract instance is not approved, a new refined one may be proposed, expressing a new intent. If a contract is approved, it can be \textbf{executed} to complete the dataflow, if all conditions are met. These three operations together manage the lifecyle of each contract, from its creation until its completion.

\begin{myitemize}

\item \textsf{propose}. Creates a new contract that specifies destination and source agents, data elements, $f()$, pre-, and postconditions.

\item \textsf{approve}. Agents in $A_{src}$ for a given contract can approve it.

\item \textsf{execute}. When a contract is approved by all agents in $A_{src}$, it can be executed. If all preconditions and postconditions are met, executing a contract gives the destination agents access to $f(DE)$, completing a dataflow.

\end{myitemize}

\mypar{Contract operations help agents work towards the sharing goal} As stated in Section~\ref{sec:model}, agents do not know each others' goal and constraint states at the beginning. When an agent proposes a contract, their proposal communicates a step towards their goal state that other agents otherwise would not have known. The other source agents will approve the contract only if it does not violate their constraints. And their approval (or the lack of it) also communicates part of their constraint states. Therefore, as agents propose and approve contracts, they learn about each others' goal and constraints, and after each execution agents move closer to a common goal state while avoiding the constraint states.

We illustrate how agents use the contract abstraction to enable sharing in the fraud detection example with two banks in \F\ref{fig:contract}. At time 0, bank 1 has the goal state of accessing a model trained over both banks’ joint data. To reach this goal, bank 1 proposes a contract that specifies: destination agents are both banks, DEs are both banks’ data, and $f()$ is the model training function. If bank 2 approves this contract, bank 1 can execute it and achieve their goal. However, this contract does not capture bank 2’s constraints
because bank 2 requires that every bank must contribute sufficient data (precondition), and that the result model must meet an accuracy threshold (postcondition). Thus, bank 2 does not approve the original contract, but instead proposes a new one with these conditions added. Bank 1 reviews the new contract and decides to approve it because the new changes do not violate bank 1’s constraints. The contract is now ready for execution, and both banks have the guarantee that the model will not be released unless all conditions are met. Due to the banks' usage of contracts, no other dataflows have occurred, and the only dataflow is one that has entered a goal state and not violated any constraint states.

\mypar{The contract abstraction can form a legally binding contract} In law, a contract is an enforceable promise between two or more parties. The following definitions are legal terminology that are different from those used in the rest of the paper. Contracts require four elements. 

\begin{myitemize}
\item \textbf{Offer}: Proposal that indicates a willingness to be bound on the specified terms.
\item \textbf{Acceptance}: Agreement to the terms of the offer.
\item \textbf{Consideration}: Value exchanged by parties.
\item \textbf{Manifestation of Mutual Assent}: Objective indication that parties intend to form an agreement.

\end{myitemize}

If any element is missing, no contract is formed. Contracts can be written, oral, or implied in actions, and they can be expressed in different forms and languages, including via programming code, as long as they fulfill the elements above.


Therefore, the contract abstraction could be designed to form a legally enforceable contract if the parties so choose. \emph{Offer}, \textit{Acceptance}, and a \textit{Manifestation of Mutual Assent} follow from the contract abstraction definition, and each party’s data and any promised dataflows constitute \textit{Consideration}.

This is an important characteristic of the abstraction because it is used to solve data sharing problems that include sensitive data. Legal contracts are often subject to cumbersome data sharing and use agreements, so parties may wish to supplement or displace them with the simplified contracts that the abstraction facilitates.

\subsection{Delegated Contract Execution}
\label{subsec:delegated}

\begin{figure}%
    \centering
    \includegraphics[width=\columnwidth]{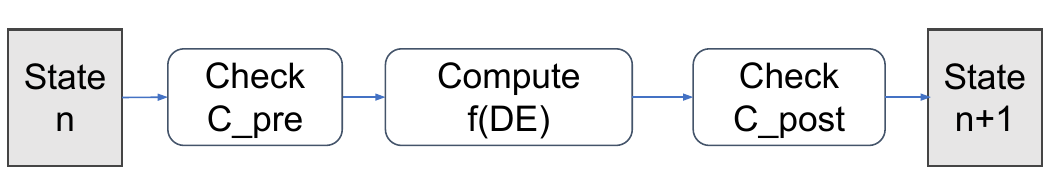}
    \caption{Sharing states separated by the contract execution}%
    \label{fig:dataflow}%
\end{figure}

Once it is approved, a contract needs to be executed. We now detail the steps to execute such a contract. In \F~\ref{fig:dataflow}, two consecutive sharing states are separated by the execution of a contract. During contract execution: i) preconditions are checked, ii) $f(DE)$ is computed, and iii) postconditions are checked. If conditions are met, the dataflow materializes. The key challenge is to ensure that $f$ runs on $DE$ to produce a result conforming to the conditions, \emph{without} agents seeing the original data. If a participating agent executes the contract, that agent would gain access to all the contract data elements, violating the contract's promise that destination agents should only gain access to $f(DE)$, not $DE$.

Instead, contract execution must be \emph{delegated} to an entity that is not a participating agent, who can ensure that the terms of the contract are implemented. There are several options:

\begin{myitemize}
\item Participants may delegate their computation to a \textit{trusted} third-party (e.g., a consultant) \cite{ilanronco}, and/or rely on a functioning legal system for enforcing the terms of the contract \cite{sjostrom2005due}.
\item Contract execution may be delegated to a \textit{trustworthy} intermediary built with security and cryptographic techniques, such as data escrows \cite{xia2023data, russinovich2019ccf}, secure multi-party computation \cite{bater2017smcql, zheng2021cerebro}, homomorphic encryption \cite{lattigo}, etc.
\item Contracts may be executed without the intervention of a legal system or a third-party by using blockchain technology \cite{amiri2021qanaat}.\footnote{A smart contract could be developed to implement the contract abstraction we propose here.}
\end{myitemize}


Using the contract abstraction in practice requires a solution for its execution, but the abstraction itself is orthogonal to the specific solution used. In the next section, we concentrate on how to program data sharing applications that use the contract abstraction and permit delegating its execution.

\section{The Contract Programming Model}
\label{sec:api}

We implement a contract programming model that permits developers to write data-sharing applications using contracts. We first enumerate five desiderata for our implementation, and then present technical contributions that satisfy those five points.

\mypar{D1: Programming Data Sharing Applications} Allow agents to program a wide range of data sharing problems, i.e., supporting arbitrary number of agents, functions, and data elements. (Section~\ref{subsec:r1}).


\mypar{D2: Contract APIs} Provide APIs to manage the contract lifecycles (i.e. proposal, approval, and execution) (Section~\ref{subsec:r2}).

\mypar{D3: Maintain Computational Performance} Provide mechanisms to mitigate the computational inefficiencies introduced by contract management (Section~\ref{subsec:r3}).

\mypar{D4: Improve Human Efficiency} Minimize human intervention by including humans (agents) only when strictly necessary to control dataflows (Section~\ref{subsec:r4}).

\noindent\textbf{D5: Support Delegated Contract Execution} without permitting unintended dataflows (Section~\ref{subsec:r5}).


We present the technical contributions to address \textbf{D1-4} in isolation from the computing infrastructure and turn our attention to this when addressing \textbf{D5}. 

\subsection{Programming Data Sharing Applications}
\label{subsec:r1}

A \emph{developer} writes a program\footnote{We use Python in our prototype, but the CPM can be implemented in other languages} that implements: i) the endpoints accessible to the intended participating \emph{agents} of the data sharing application; ii) the functions ($f()$) available to participating agents. The developer uses the following annotations:


\mypar{\textsf{@api\_endpoint}} This makes the annotated function publicly available to participating agents. The developer must choose carefully which functions to expose.


\mypar{\textsf{@contract\_function}} This indicates the annotated function accesses data elements and thus requires a contract; they represent the $f()$ in the contract abstraction. Functions annotated with \textsf{@contract\_function} may also be annotated with \textsf{@api\_endpoint}. The pre- and post-conditions are implemented programmatically within these \textsf{@contract\_function}-annotated functions, thus making full use of the programming language's conditional logic and ability to compute arbitrary functions.

Later in \F~\ref{fig:code_three} from Section \ref{subsec:rq2}, the annotations are used in the fraud detection example. 
In the program, \textsf{upload\_credit\_transaction\\\_data} is annotated with \textsf{@api\_endpoint}, and banks call this to upload their data. Meanwhile, \textsf{show\_schema} is annotated with both \textsf{@api\_endpoint} and \textsf{@contract\_function}. Banks call this to see the schema of every data element in the contract.

Any function may access external software libraries, and a CPM program may contain functions without any annotations, thus permitting developers to factor out common tasks and follow standard software engineering practices to write more complex applications.


The CPM program is executed by an entity other than the participating agents, following the principles of delegated computation outlined in Section \ref{subsec:delegated}. It is a technical solution implemented on a third-party server, which we explain in more detail in Section \ref{subsec:r5}.



\mypar{Deployment} \textsf{@api\_endpoint}-annotated functions are available to participating agents through an implementation-dependent interface. Our implementation exposes such functions via a REST-API. A server hosts the sharing application, and participating agents connect to the server via a client to call the \textsf{@api\_endpoints} and \textsf{@contract\_functions}.

\subsection{Contract APIs}
\label{subsec:r2}

The CPM provides developers with a \textsf{contract\_api} library to create and manage contracts.

\mypar{\textsf{register\_agent(id, name)}} Registers an agent (using, optionally, an externally assigned ID such as by an organization IAM's system) in the sharing application so its ID is available to other agents. Agent IDs are used to address agents as source and destination agents when creating contracts.

\mypar{\textsf{DEStore}} A \textsf{DEStore} permits: i) registering (writing) data elements and assigning them an ID so they can be referred to during contract creation; ii) reading previously registered data elements. There are different implementations catering to different data element formats. For example, we have implemented a \textsf{FileStorageDEStore()} that provides access to the file system to handle files in various formats (e.g. CSV and Parquet). 


\mypar{\textsf{propose\_contract(•), approve\_contract(•), deny\_contract(•), get\_pending\_contracts(agent\_id)}} The \textsf{contract\_api} contains these functions (arguments redacted because they match those of the contract abstraction), which developers can expose to participating agents through a \textsf{@api\_endpoint}-annotated function.


Notably, \textsf{contract\_api} does not provide a function to \emph{execute} contracts. Agents call \textsf{@contract\_function}-annotated functions in the sharing program and, if there is a contract available that permits the dataflows created by such functions, then these execute.

\subsection{Maintain Computational Performance}
\label{subsec:r3}

Programming applications with contracts affects performance in two ways. First, it complicates the program's state management. Second, it potentially introduces unnecessary computation. We augment our implementation of the CPM to address both.


\subsubsection{Program State Management} Only the final result of a contract can be revealed, so developers sometimes need write data sharing applications that are less efficient than those written without contracts. Consider a group of agents who want to run function $f$, which first combines data from different agents, and then trains a model over the combined data. $f$ can be decomposed into the sequential execution of two functions: $g$, which combines the data, and $h$, which trains the model. $h(g)$ is equivalent to $f$. But, if agents want to train a different model in $f$, $f$ still needs to execute $g$ to combine the data from scratch, even though the result $g$ will not change. This introduces inefficiencies. Unfortunately, if the developer instead writes an application that separates $g$ and $h$ into two functions, then a contract over $g$ will disclose the combined data to the participating agents, which may violate agents' constraints.

The solution is to allow developers to write \textit{intermediate data elements} that are computed during the execution of a function, but will not become accessible to any destination agent, i.e., the destination agents are empty. Developers write the intermediate data elements how they would write any other data element: with \textsf{DEStores}. Internally, if a \textsf{DEStore.write()} is called during the execution of a \textsf{@contract\_function}, the system will track all data elements accessed by the function as provenance for this intermediate data element. This provenance is used to identify which source agents need to be incorporated into a contract that involves accessing intermediate data elements. Reusing intermediate \textit{effectively} decouples the execution of a \textsf{@contract\_function}, without actually separating it into different functions. The output remains the same, but the potentially large performance impact is avoided. We demonstrate in the evaluation (Section \ref{sec:evaluation}) how this optimization contributes to the performant execution of data sharing applications.

\subsubsection{Short-circuiting Function Execution} 

Developers can write arbitrary \textsf{@contract\_function} in the sharing program, including ones that either maliciously or unintentionally access data outside of the approved contract. Our system  ensures that a function finishes execution only if it has solely used approved data elements by its contract; otherwise, the result will not be released. Detecting these illegal data accesses early in function execution can lead to performance improvements, and we introduce a short-circuiting technique to skip computation whenever possible.

The data elements accessed so far in the execution are tracked by calls to \textsf{DEStore}, and they are continuously compared with the data elements that a \textsf{contract\_function} is permitted to access from contracts. Whenever the data elements accessed do not belong in an approved contract, the environment suspends (i.e., short-circuits) the function execution, skipping computation that would otherwise be wasted. The environment tracks the accesses closely and therefore introduces a slight overhead, but it is insignificant in comparison to the large gains of skipping computation. We demonstrate this in the evaluation section.

\subsection{Improve Human Efficiency}
\label{subsec:r4}

Data sharing requires human (agent) intervention. Source agents need to closely inspect proposed contracts and discuss them with compliance (and other relevant departments) before carefully choosing an action. However, there are situations where agents propose the same contract repeatedly. In those cases, human intervention could be saved if the approval process is codified ahead of the contract proposal. We introduce \emph{contract management rules} to minimize agent intervention for such cases.



\mypar{Contract Management Rule} Source agents may know their constraints on what contracts they will approve, potentially even without an active contract proposal. For such instances, they can express these through a contract management rule using a \textsf{register\_cmr(•)} function in the \textsf{contract\_api}. When another agent proposes a contract involving these source agents, such a contract is first checked against existing contract management rules. If the rule permits the contract, then it is approved without human intervention. If not, then human is involved and the contract lifecycle is started.

A contract management rule is related to access control mechanisms, such as role-based access control and attribute-based access control. Information conveyed in a rule automatically specifies a set of contracts to be approved or rejected. Contract management rules greatly reduce human intervention in many sharing problems. In the fraud detection example, one \textsf{@contract\_function} will return the schema of all contract data elements. If one bank does not consider this information as sensitive, they can specify a rule that automatically approves any contracts that execute this function, saving them from the need of manually approving each one. 

\subsection{Support Delegated Contract Execution}
\label{subsec:r5}




Of all solutions that can be used for delegated contract execution, we prefer one that does not require \emph{trusting} other agents to not leak data. For example, one option mentioned in Section \ref{subsec:delegated} is for agents to delegate their computation to a consultant---indeed, this is largely how these data-sharing consortia are set up today. Yet, relying on legal safeguards is both cost-intensive and slow, and agents are still prone to others not enforcing the terms of the contract.

We choose to implement the CPM on top of a data escrow~\cite{xia2023data} because of the following advantages it offers:

\begin{myitemize}
\item \textbf{R1.} Data is encrypted end-to-end.
\item \textbf{R2.} Without notice, the implementation of the \textsf{contract\_api} cannot be tampered with, e.g., installing a backdoor that leaks data.
\item \textbf{R3.} Contracts cannot be fabricated to permit dataflows that no agent have approved without notice.
\end{myitemize}

These features of a data escrow architecture makes it suitable for implementing delegated contract execution. Data elements are only accessible to the escrow by default, and they can only be accessed through previously and explicitly approved computation delegated from the agent querying the data. In \textsc{programmable dataflows}, contracts specify the conditions that establish whether computation can run over data elements.


\mypar{Implementation}
There can be multiple ways to implement a data escrow architecture that meet R1-3. Our data escrow implementation follows that proposed in ~\cite{xia2023data}, which relies on hardware-based trusted execution environments and cryptographic techniques to ensure that data is encrypted end-to-end, including during computation. (\textbf{R1}) The software is remotely attested so any modifications are noticed. (\textbf{R2}) The contract-related state is stored as part of the escrow state, which is an in-memory database---and thus protected by the same mechanisms that protect data elements. (\textbf{R3}) The escrow is the server hosting the data sharing program and thus enabling delegated contract execution.

\section{Evaluation}
\label{sec:evaluation}

We design the evaluation to address the main hypotheses of the work, listed below as research questions:

\begin{myitemize}

\item \textbf{RQ1:} Can we model varied data sharing problems with the contract abstraction? (Section~\ref{subsec:rq1}).

\item \textbf{RQ2:} Is it easier to implement sharing problems using the CPM than using alternative data sharing technologies? (Section~\ref{subsec:rq2})

\item \textbf{RQ3:} Do the programming model extensions permit writing efficient data sharing programs? (Section~\ref{subsec:rq3})

\end{myitemize}


\begin{table*}[ht]
\small
    \centering
    \begin{tabular}{| P{0.16\textwidth} | P{0.30\linewidth} | P{0.24\linewidth} | P{0.21\linewidth} |}
        \hline
        \textbf{Sharing Problem} & \textbf{Agents' Goals} & \textbf{Agents' Constraints} & \textbf{Illustrating Patterns of Dataflow(s)} \\
        \hline
        \hline
        Healthcare federation \cite{bater2017smcql, bater2020saqe, bater2018shrinkwrap, he2017composing}
        & \raggedright
        \vspace{0.4\baselineskip}
        Healthcare institutions jointly analyze patient data. (e.g. identify common patients between hospitals)
        \vspace{0.4\baselineskip}
        &  \raggedright
        Cannot share raw patient data \cite{bater2017smcql, bater2020saqe, bater2018shrinkwrap, he2017composing}
        & 
        many-to-many
        \multirow{3}{*}{\includegraphics[width=\linewidth]{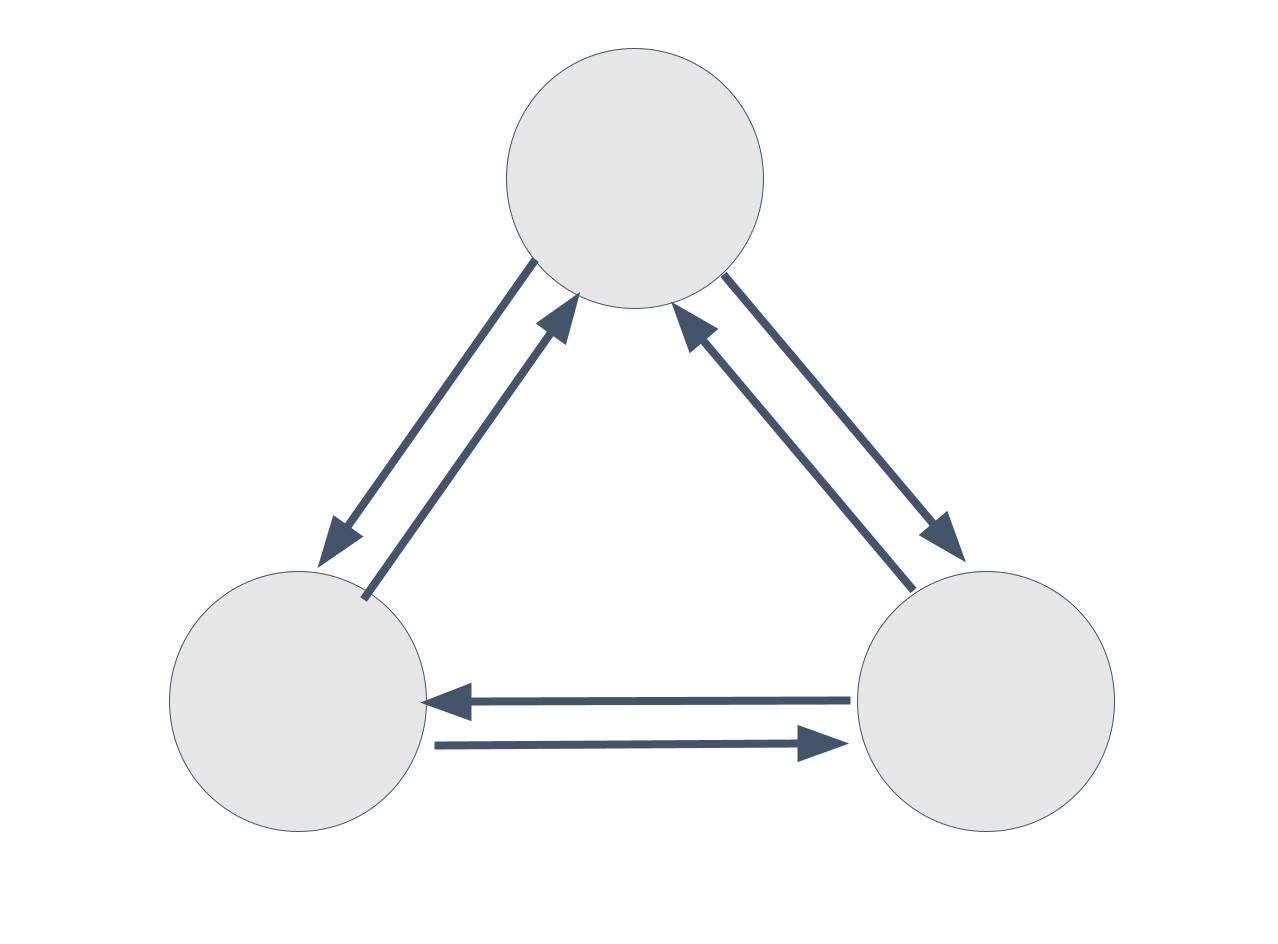}}
        \tabularnewline
        \cline{1-3}
        Mobile network federation \cite{tong2022hu}
        & \raggedright
        \vspace{0.4\baselineskip}
        Group of mobile network providers cooperate to identify user's travel history.
        \vspace{0.4\baselineskip}
        &  \raggedright
        Cannot share raw customer data \cite{tong2022hu}
        & 
        \tabularnewline
        \cline{1-3}
        Fire department federation \cite{clifton2004privacy}
        & \raggedright
        \vspace{0.4\baselineskip}
        Fire departments share data between each other to develop new training programs.
        \vspace{0.4\baselineskip}
        &  \raggedright
        Prevent departments' identities to be linked to shared data \cite{clifton2004privacy}
        & 
        \tabularnewline
        \hline
        Online research data sharing \cite{johnson2016mimic}
        & \raggedright
        \vspace{0.9\baselineskip}
        Institutions share data online for researchers/scientists to access.
        \vspace{0.9\baselineskip}
        &  \raggedright
        Only agents who have completed a training can download the data. \cite{johnson2016mimic}
        & 
        one-to-many
        \multirow{2}{*}{\includegraphics[width=0.8\linewidth]{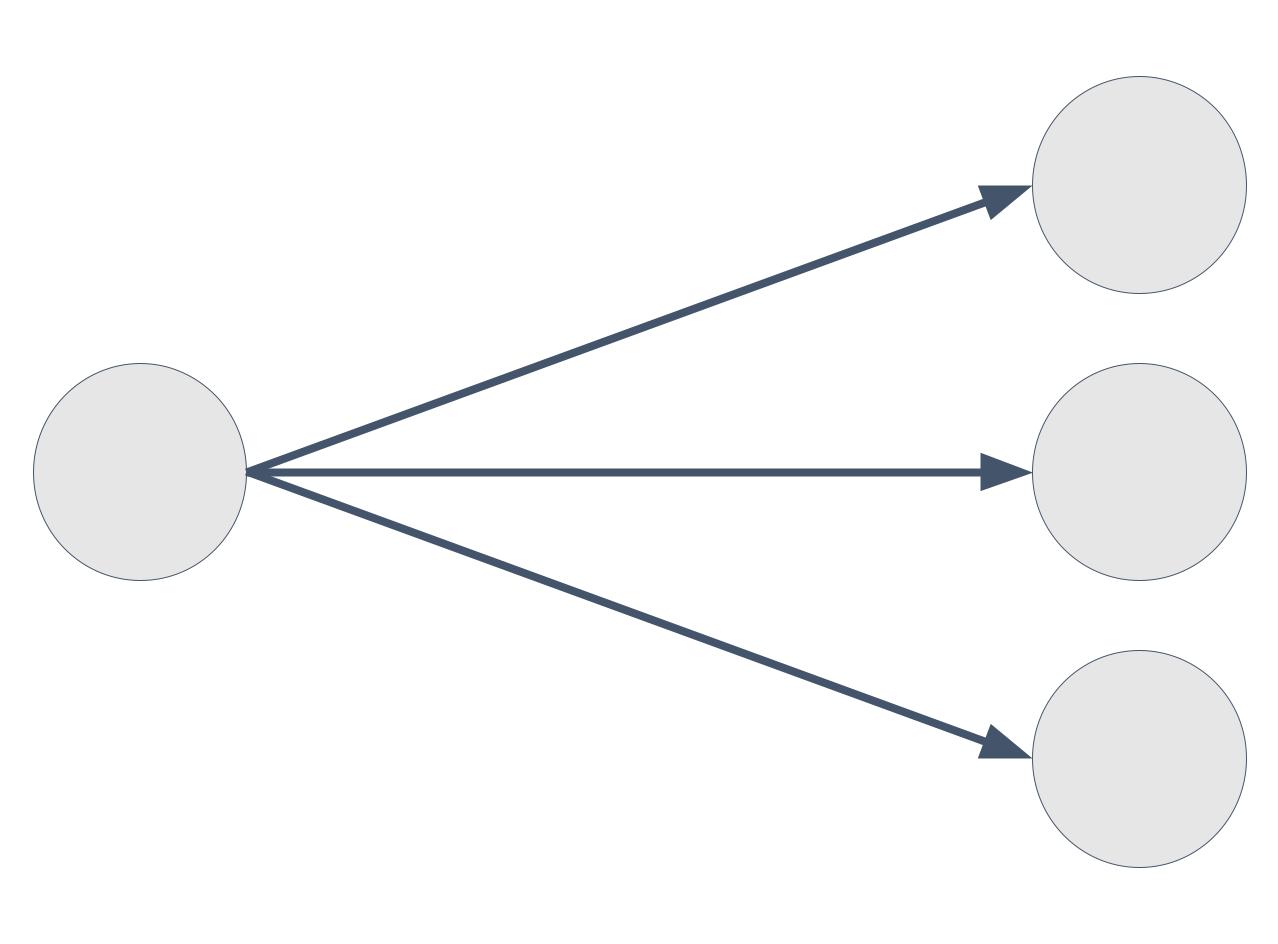}}
        \vspace{1.2\baselineskip}
        \tabularnewline
        \cline{1-3}
        Supply chain data sharing
        & \raggedright
        \vspace{0.75\baselineskip}
        Retailers and manufacturers share data about products and sales with business partners 
        \cite{databricks_2022}.
        \vspace{0.75\baselineskip}
        &  \raggedright
        Only selected partners get access to data
        & 
        \tabularnewline
        \hline
        Crowdsourcing platforms 
        \cite{mturk}
        & \raggedright
        \vspace{0.2\baselineskip}
        Data requesters pay workers to contribute certain data (e.g. transcribe audio recordings).
        \vspace{0.2\baselineskip}
        &  \raggedright
        Workers need payment; requesters specify who are eligible workers
        & 
        one-to-one
        \multirow{3}{*}{
        \includegraphics[width=0.8\linewidth]{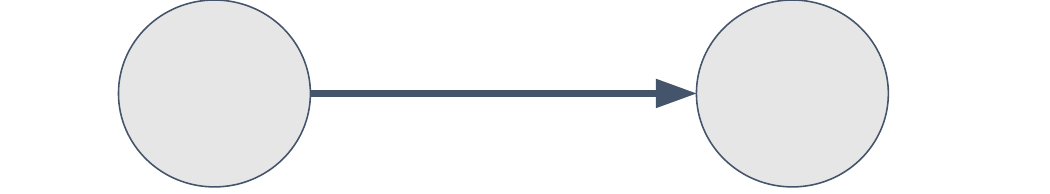}}
        \tabularnewline
        \cline{1-3}
        Data markets \cite{awsmarket, snowflakemarket}
        & \raggedright
        Buyers purchase datasets from providers.
        &  \raggedright
        Sellers need payment; buyers want data to meet their needs
        & 
        \tabularnewline
        \hline
        Training Google's keyboard prediction algorithm 
        \cite{hard2018federated}
        & \raggedright
        \vspace{0.9\baselineskip}
        Google collects user-generated text to train a keyboard prediction algorithm.
        \vspace{0.9\baselineskip}
        &  \raggedright
        Users don't want Google to save their generated text;
        & 
        many-to-one
        \multirow{2}{*}{\includegraphics[width=0.75\linewidth]{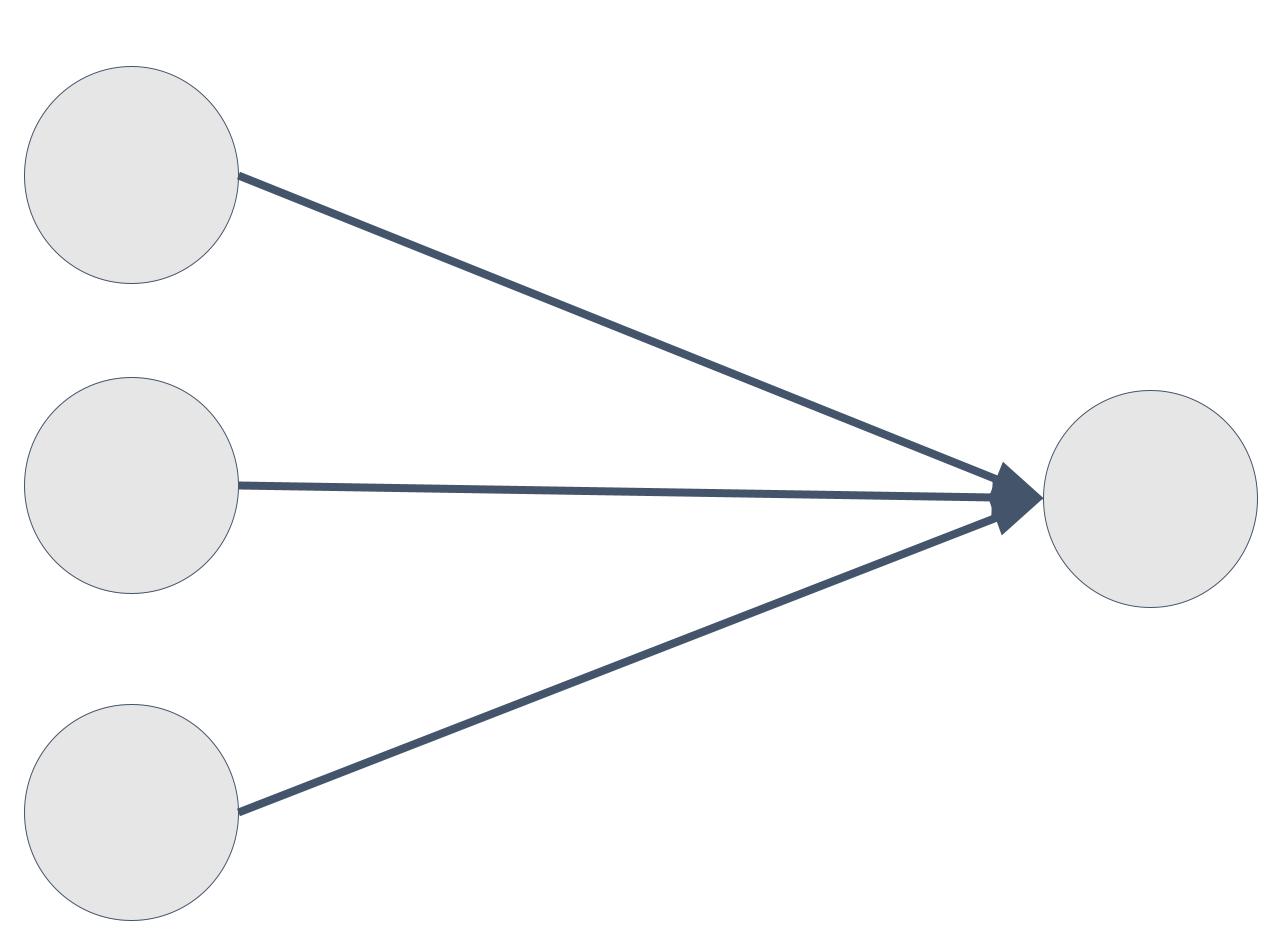}}
        \vspace{1\baselineskip}
        \tabularnewline
        \cline{1-3}
        Chrome collecting users' homepage distribution \cite{erlingsson2014rappor}
        & \raggedright
        \vspace{0.9\baselineskip}
        Chrome collects homepage distribution to improve service security.
        \vspace{0.9\baselineskip}
        &  \raggedright
        Users with rare homepage settings don't want to disclose it
        & 
        \tabularnewline
        \hline
        US Census
        & \raggedright
        The census bureau collects data from all citizens, then publishes statistics about collected data.
        &  \raggedright
        The census bureau cannot publish information that identifies individuals
        & \raggedright
        many-to-one; one-to-many
        \tabularnewline
        \hline
        Training LLM from web data
        & \raggedright
        \vspace{0.2\baselineskip}
        An entity collects data published by individuals on the web over time to train an LLM.
        \vspace{0.2\baselineskip}
        &  \raggedright
        \vspace{0.2\baselineskip}
        Individuals may not want their data to be used for training
        \vspace{0.2\baselineskip}
        & \raggedright
        one-to-many; many-to-one
        \tabularnewline
        \hline
    \end{tabular}
    \caption{Collection of dataflows that arise in various data sharing problems.} 
    \label{tab:dataflows_contracts}

\end{table*}



\subsection{RQ1: Expressiveness of Contract Abstraction}
\label{subsec:rq1}


We show the contract abstraction applies to a diversity of data sharing problems. We compile data sharing problems from descriptions of real deployments and from the literature (see column "Sharing Problem") and show them as rows in the Table~\ref{tab:dataflows_contracts}.

For each problem, the table includes: i) agent's goals, ii) agents' constraints, and iii) a diagram of dataflow(s) that take place in the sharing problem, where nodes represent agents, and (directed) edges represent dataflows.


We classify the sharing problems into a set of \emph{canonical} problems described by their dataflow patterns. Qualitatively, we show that the contract abstraction can represent the underlying dataflows in all of the patterns.

\mypar{Many-to-many pattern} In this pattern, a set of agents want to run computation over their joint data. Every agent contributes data and observes the results of the computation. The corresponding contract includes all agents in the federation as the source and destination agents. For example, consider the mobile network federation \cite{tong2022hu}. Let the providers in the federation be $A_{\text{net}}$ and their data $D_{\text{net}}$. Then, a contract to enable this federation has \(\langle\)$A_{\text{dest}}$ = $A_{\text{src}}$ = \{$A_{\text{net}}$\}, $DE$ = \{$D_{\text{net}}$\}, $f$ = \{spatial queries\}\(\rangle\). Many sharing problems with this dataflow pattern have the constraint that agents cannot share the raw data with each other. The contract enforces this by construction, as the destination agents would only get access to $f(DE)$.

\mypar{One-to-many pattern} In this pattern, one source agent wants to share data with multiple destination agents, provided they meet certain conditions. For example, an online medical database \cite{johnson2016mimic} (MIMIC-II, denoted $a_{\text{m}}$), requires users $A_{\text{u}}$ to have completed a training program before accessing the database (denoted $d_{\text{m}}$). Let the requirement for providing training completion be denoted $c_{\text{train}}$. Then a contract that meets MIMIC-II's requirements is as follows: \(\langle\)$A_{\text{dest}}$ = \{$A_{\text{u}}$\}, $A_{\text{src}}$ = \{$a_{\text{m}}$\}, $DE$ = \{$d_{\text{m}}$\}, $f$ = \{direct share\}, $C_{\text{pre}}$ = \ \{$c_{\text{train}}$\}\(\rangle\). Implementation-wise, $c_{\text{train}}$ can be encoded with a function that takes in proof of training completion, then returns the data.

\mypar{One-to-one pattern} In this pattern, a destination agent wants to access data that meet certain requirements. They can choose from a pool of potential source agents, who want to get something in exchange for their data. This pattern is common in crowdsourcing platforms \cite{mturk} and online data marketplaces \cite{awsmarket, snowflakemarket}. For example, an e-commerce company is looking for training data that can improve the accuracy of their recommendation system by at least ten percent on some test data. An important challenge faced by agents in these problems is the \textit{Arrow information paradox} \cite{arrow}: destination agents are not ready to commit to a payment before knowing the value of data, but source agents will not grant access without payment. The contract abstraction helps by incorporating buyers' requirements, the conditions under which they would pay, into the contract's conditions. For example, in a data market for ML, a buyer requires that a private model augmented with the seller's data should achieve a performance increase by three percent (denoted $c_{\text{perf}}$). The following contract captures this requirement: \(\langle\)$A_{\text{dest}}$ = \{$a_{\text{buyer}}$\}, $A_{\text{src}}$ = \{$a_{\text{seller}}$\}, $DE$ = \{$d_{\text{seller}}$\}, $f$ = \{direct share\}, $C_{\text{pre}}$ = \ \{$c_{\text{perf}}$\}\(\rangle\).

\mypar{Many-to-one pattern} In this pattern, one destination agent wants to collect data from many source agents. One example is Google training its keyboard prediction algorithm from user-generated text \cite{hard2018federated}. Today, source agents have little control over their data: once their data is collected, they are "at the mercy" of how the destination agent will use their data. The contract abstraction helps the source agents gain more control over their data by allowing them to specify their constraints explicitly in a contract. For example, denote Google by $a_g$, the users from which Google collect data by $A_u$, users' data by $D_u$, and the keyboard prediction algorithm by $f_{\text{key}}$. Then, the following contract ensures that Google only gets the model trained from the users' data, but not their raw data: $A_{\text{dest}}$ = \{$a_g$\}, $A_{\text{src}}$ = \{$A_u$\}, $DE$ = \{$D_u$\}, and $f$ = \{$f_{\text{key}}$\}.

\mypar{Mix of patterns} Other sharing problems exhibit multiple patterns. We take the government census as an example. At a high level, the census consists of two sequential patterns: a many-to-one followed by a one-to-many. The bureau first collects data from every citizen, then publishes aggregate statistics about collected data. There exists one challenge: the census bureau is subject to law \cite{census_code} that they cannot publish information that identifies individuals. Yet, after the data collection period, the bureau already has every citizen's original data, and citizens have no way of ensuring, technically, that bureau will keep their data confidential. In other words, the citizens have already lost control over their data, once their raw data is shared, and this has historically been a source of concern for citizens. A contract remedies this situation by only giving access to $f(DE)$, where $f()$ can be a function that adds noise to individual's data, or a function that aggregates multiple agents' data together. 

\mypar{Summary of Qualitative Results} We show the contract abstraction represents all the canonical patterns identified, and that such canonical patterns express multiple data sharing problems.


\subsection{RQ2: Effectiveness of the CPM}
\label{subsec:rq2}


To study if the CPM is effective for implementing data sharing problems, we introduce three data sharing problems in detail: \textit{healthcare data sharing} \cite{schmidt2015danish}, \textit{ad matching} \cite{iab_report}, and \textit{financial fraud detection} from the running example. These examples reflect the complexity and constraints of many sharing problems, they encompass multiple dataflows and comprise a diverse set of goals and constraints. We show how the desiderata from Section 5 arise from these three examples and analyze their implementation using programmable dataflows. To highlight the qualities of the implementation, we compare it qualitatively with alternative sharing baselines. We first describe the three sharing problems.


\begin{table*}[ht]
\small
    \centering
    \begin{tabular}{| P{0.2\textwidth} | P{0.08\linewidth} | P{0.08\linewidth} | P{0.08\linewidth} | P{0.08\linewidth} | P{0.12\linewidth} | 
    P{0.1\linewidth} | P{0.08\linewidth} |}
        \hline
         & Inner Product & Set Intersection & Pearson Correlation & Ridge Regression & AES & Logistic Regression & Sorting \\
        \hline
        \hline
        Generic Framework (ABY, Moose, MPyC) & 7 & >200 & 7 & >100 & \textasciitilde800 with ABY, \textasciitilde100 with MPyC & \textasciitilde50 with PyMoose & 8 with MPyC\\
        \hline
        Using Python Library/Native Python & 1 & 1 & 2 & <5 & <10 & <5 & 1 \\
        \hline
    \end{tabular}
    \caption{LoC Needed to Implement Certain f() Using Generic Secure Frameworks}
    \label{tab:loc}

\end{table*}

\begin{table}[ht]

\small
    \centering
    \begin{tabular}{| P{0.22\columnwidth} | P{0.22\columnwidth} | P{0.18\columnwidth} | P{0.20\columnwidth} |}
        \hline
        & \textbf{Support Arbitrary Computation ($f$)} & \textbf{Scalable to Multiple Sources (3+)} & \textbf{Support Preconditions \& Postconditions}  \\
        \hline
        \hline
        SMCQL \cite{bater2017smcql}  & \XSolidBrush \ (SQL) & \XSolidBrush \ (2) & \XSolidBrush  \\
        \hline
        Shrinkwrap \cite{bater2018shrinkwrap}  & \XSolidBrush \ (SQL) & \XSolidBrush \ (2) & \XSolidBrush  \\
        \hline
        SAQE \cite{bater2020saqe}  & \XSolidBrush \ (SQL) & \XSolidBrush \ (2) & \XSolidBrush  \\
        \hline
        Conclave \cite{volgushev2019conclave}  & \XSolidBrush \ (SQL) & \XSolidBrush \ (3) & \XSolidBrush  \\
        \hline
        Secure Yannakakis \cite{wang2021secure}  & \XSolidBrush \ (SQL) & \XSolidBrush \ (2) & \XSolidBrush  \\
        \hline
        Senate \cite{poddar2021senate}  & \XSolidBrush \ (SQL) & \Checkmark & \XSolidBrush  \\
        \hline
        Snowflake Cleanroom \cite{kaufman_2022}  & \XSolidBrush \ (SQL) & \XSolidBrush \ (2) & \XSolidBrush  \\
        \hline
        Federated Learning \cite{ziller2021pysyft, beutel2020flower} & \XSolidBrush \ (ML) & \Checkmark & \XSolidBrush \\
        \hline
        Cerebro \cite{zheng2021cerebro} & \XSolidBrush \ (ML) & \Checkmark & \Checkmark \\
        \hline
        Hu-Fu \cite{tong2022hu} & \XSolidBrush \ (spatial queries) & \Checkmark & \XSolidBrush \\
        \hline
        PSI \cite{de2010practical, chen2017fast, hazay2017scalable, kolesnikov2017practical, li2021prism} & \XSolidBrush \ (set intersection) & \Checkmark \& \XSolidBrush * & \XSolidBrush \\
        \hline
        Private Record Linkage \cite{he2017composing} & \XSolidBrush \ (record linkage) & \XSolidBrush \ (2) & \XSolidBrush \\
        \hline
    \end{tabular}
    \caption{The Space of Supported Dataflows by a Collection of Existing Data Sharing Technologies.}
    \label{tab:technologies}

\end{table}

\vspace{-1mm}

\mypar{Healthcare Data Sharing} The Danish National Patient Registry (DNPR) contains medical data about Danish citizens since 1977, which may provide immense value to researchers. For instance, the data may be used in causal inference to \emph{``obtain information on confounders, particularly co-morbidities''} \cite{schmidt2015danish}. As an example, the Danish Adult Diabetes Registry (DADR) contains data on patients’ smoking status and HbA1c (measured blood sugar level). They want to calculate the causal effect of smoking (the treatment variable T) on HbA1c (the outcome variable O), and they know that depression is a confounder for these two variables. DNPR has data on patients' historical record of depression, and having access to this data can help DADR get more accurate estimate of the causal effect. While allowing access to DNPR's data is clearly beneficial, DNPR is subject to the Danish Data Protection Act \cite{danish_data_protection}, which specifies that their data can only be processed for scientific purposes. So DNPR cannot directly send their data to DADR, or make their data public, because there is no guarantee that the above constraint will be satisfied. 

This sharing problem has three requirements. i) DNPR needs to ensure that their data is only processed for scientific purposes. Thus, they only allow one type of computation that joins agent data with additional confounders (i.e. columns) from DNPR's data, then calculates the causal effect between a treatment variable and an outcome variable. ii) Agents' data should include the attribute "CPR", which is a join key that uniquely identifies each Danish citizen \cite{schmidt2015danish}, so that their data can join with DNPR's. iii) A large number of agents want to use DNPR's data, so DNPR wants to automatically approve all those requests to avoid human intervention.

\mypar{Ad Matching} Advertisers want to run analysis over data collected by media publishers to study the effect of their ad campaigns~\cite{iab_report}. Don, an advertiser, wants to study the click-through rate of an ad over data collected by Facebook and Youtube, which jointly includes audiences' responses to an ad, their demographics, and their social media interactions. Don has some idea of what analysis are helpful, such as identity matching (return overlapping users across Facebook and Youtube's data), user profile expansion (join information about overlapping users), user discovery (identify groups of users with certain characteristics), and machine learning (identify important factors that influence users' click-through rate). But Don cannot specify all the computations he will run at time 0 (e.g. what queries to ask, what types of model to train, etc.), because such analysis is exploratory in nature. Meanwhile, Don knows that the media publishers will not reveal their plaintext data, but certain computations on the data (e.g. aggregate analysis) may be allowed.

This sharing problem has two requirements. i) Due to the exploratory nature of Don's analysis, he wants functionalities to run a wide range of $f()$, ranging from analytical queries to ML models. ii) Don's analysis should be approved on-the-fly because Don does not know Facebook and YouTube's constraints at time 0, nor do Facebook and YouTube know what analysis Don wants to run.

\mypar{Financial Fraud Detection} We revisit the example from the Introduction. This sharing problem has three requirements. i) At time 0, banks do not know if their data is readily compatible, nor do they know what other banks are willing to disclose about their data. Thus, the banks want support for data preparation functions that disclose information about input data (e.g. show sample, show schema, etc.). ii) Banks not only need functionality to train a model over their joint data, but they also need to incorporate a precondition that each bank contributes enough training data, and a postcondition that the result model will only be released upon achieving some minimum accuracy on the combined test set. iii) Similar to ad matching, banks want to approve each data access on-the-fly.

\mypar{Summary of Agents' Requirements} We summarize the agents' requirements across the three sharing problems as follows:

\noindent\emph{1. Support a wide range of dataflows.} The $f()$ agents want to run in the three problems include SQL, causal queries, and training ML models with customized conditions. The cardinality of agents also differ in each problem.

\noindent\emph{2. Support data preparation.} Without knowledge over other agents' data elements, agents want to ensure that their data is compatible for the desire computation to run. In healthcare data sharing, DNPR wants to ensure that other agents' data can be joined with their data. In fraud detection, the banks want to ensure that their data is combined properly before a joint model is trained.

\noindent\emph{3. Ensure computational efficiency.} In ad matching, Don may need to train multiple models over the media publisher's joint data for comparison, due to the exploratory nature of his analysis. He wants to do so efficiently.

\noindent\emph{4. Effectively manage contract lifecycles.} In the ad matching and fraud detection, agents have to propose and approve contracts on-the-fly, because they do not have complete information over others' goals and constraints. On the other hand, in healthcare data sharing, DNPR wants to automatically allow all requests to run causal queries on their data, as this does not violate their constraints.

\subsubsection{Implementing the Sharing Problems Using the CPM vs Using Other Technologies} To answer the research question, we conduct a qualitative comparison of each data-sharing problem implemented using programmable dataflows and contrast that implementation with a medley of existing data sharing technologies, which we summarize in Table~\ref{tab:technologies}. Our criteria for including a technology in the table is that it has to \textit{prevent unintended dataflows}.

Throughout this section, we use a few selected code snippets (Figures \ref{fig:code_one}, \ref{fig:code_two}, and \ref{fig:code_three}) to illustrate aspects of the CPM. 

\lstset{escapeinside={<|}{|>}}
\lstset{basicstyle=\small\ttfamily}
\definecolor{mycolor}{RGB}{255, 128, 0}

\begin{figure}

    \lstset{basicstyle=\footnotesize\ttfamily}
    \centering
    \begin{lstlisting}[language=Python]
from dowhy import CausalModel

<|\textcolor{mycolor}{@api\_endpoint}|>
def upload_data_with_CPR(data):
    df = pd.read_csv(io.BytesIO(data))
    if "CPR" in df.columns:
        return CPM.FileStorageDEStore.write(data)
    else:
        return "Error: CPR column does not exist."

<|\textcolor{mycolor}{@api\_endpoint}|>
<|\textcolor{mycolor}{@contract\_function}|>
def run_causal_query(user_de_id, additional_vars, dag_spec, treatment, outcome):
    ... # augment user's data with additional confounders from DNPR's data
    model = CausalModel(data=joined_df, treatment=treatment, outcome=outcome, graph=dag_spec)
    estimate = model.estimate_effect(...) # use Python's dowhy package to calculate the causal effect
    return estimate.value

<|\textcolor{mycolor}{@api\_endpoint}|>
def upload_cmr(agent, de, f):
    return CPM.upload_cmr(agent, de, f)
...
\end{lstlisting}
\caption{Program snippet for healthcare sharing. \textsf{upload\_data\_with\_CPR} ensures all agents' data includes a "CPR" column. \textsf{run\_causal\_query} uses Python's dowhy package to calculate causal effect. \textsf{upload\_cmr} uses the default implementation from the CPM. DNPR call this to automatically allow agents to run \textsf{run\_causal\_query} over their data.}
\label{fig:code_one}
\end{figure}

\begin{figure}
    \lstset{basicstyle=\footnotesize\ttfamily}
    \centering
    \begin{lstlisting}[language=Python]
...
<|\textcolor{mycolor}{@api\_endpoint}|>
def propose_contract(dest_agents, des, f, args):
    # use CPM's default implementation to propose a contract
    return CPM.propose_contract(dest_agents, des, f, args)
<|\textcolor{mycolor}{@api\_endpoint}|>
def approve_contract(contract_id):
    # use CPM's default implementation to approve a contract
    return CPM.approve_contract(contract_id)
...
<|\textcolor{mycolor}{@api\_endpoint}|>
<|\textcolor{mycolor}{@contract\_function}|>
def train_advertising_model(model_name, label_name, query=None):
    joined_data = CPM.KVStore.read("joined_data")
    if not joined_data:
        joined_data = ... # join Facebook & YouTube's data from query
        CPM.KVStore.write(joined_data, "joined_data")
    ... # proceed to model training
\end{lstlisting}
\caption{Program snippet for ad matching. \textsf{propose\_contract} and \textsf{approve\_contract} allows contracts to be proposed and approved one at a time. \textsf{train\_advertising\_model}allows reusing the saved joined result from Facebook and YouTube's data}
\label{fig:code_two}
\end{figure}

\begin{figure}
    \lstset{basicstyle=\footnotesize\ttfamily}
    \centering
    \begin{lstlisting}[language=Python]
...
<|\textcolor{mycolor}{@api\_endpoint}|>
def upload_credit_transaction_data(data):
    '''For banks to upload their transaction data.'''
    ...
<|\textcolor{mycolor}{@api\_endpoint}|>
<|\textcolor{mycolor}{@contract\_function}|>
def show_schema(de_ids):
    '''Return schema of DEs in de_ids.'''
    ...
<|\textcolor{mycolor}{@api\_endpoint}|>
<|\textcolor{mycolor}{@contract\_function}|>
def share_sample(de_ids, sample_size):
    '''Return samples of size sample_size for DEs in de_ids.'''
    ...
<|\textcolor{mycolor}{@api\_endpoint}|>
<|\textcolor{mycolor}{@contract\_function}|>
def check_column_compatibility(de_1, de_2, cols_1, cols_2):
    '''Compare two lists of columns, each from one input DE. Computes the KS statistic for numerical columns and jaccard similarity for text columns.'''
    ...
<|\textcolor{mycolor}{@api\_endpoint}|>
<|\textcolor{mycolor}{@contract\_function}|>
def train_fraud_model(train_de_ids, size_constraint, test_de_ids, target_accuracy, label_name):
    ...
    for de_id in train_de_ids:
        de = CPM.CSVDEStore.read(de_id)
        if len(de) < size_constraint:
            return "Input size constraint failed."
    ... # more code to construct X_train, X_test, Y_train, Y_test, and to train the model
    if clf.score(X_test, Y_test) < target_accuracy:
        return "Accuracy constraint failed"
    return clf
...
\end{lstlisting}
\caption{Program snippet for fraud detection. First three \textsf{@contract\_functions} support data preparation. \textsf{train\_fraud\_model} checks the input cardinality precondition and the accuracy postcondition.}
\label{fig:code_three}
\end{figure}

\mypar{Support a wide range of dataflows}   \F\ref{fig:code_one} shows an implementation of the causal query for healthcare data sharing. It first augments user's data with additional confounders from DNPR's data, then use Python’s dowhy package to calculate a causal effect.  \F\ref{fig:code_three} shows an implementation of training the joint fraud detection model over all banks' data, while checking for both the input cardinality and accuracy conditions. Our programming model allows agents to implement customized computation across all three sharing problems using Python.

On the other hand, existing technologies are highly constrained in the dataflows they support. For example, none of the technologies in Table \ref{tab:technologies} allow developers to implement the causal query from the healthcare example easily because most only support specific functions. Some only support SQL \cite{bater2017smcql, bater2018shrinkwrap, bater2020saqe, volgushev2019conclave, wang2021secure, poddar2021senate, kaufman_2022, wang2024special}, some only ML \cite{zheng2021cerebro, ziller2021pysyft, beutel2020flower,zhang2017private}, and some only private set intersection \cite{de2010practical, chen2017fast, hazay2017scalable, kolesnikov2017practical}. In addition, many of them cannot scale to dataflows with more than three source agents \cite{bater2017smcql, bater2018shrinkwrap, bater2020saqe, volgushev2019conclave, wang2021secure, kaufman_2022, he2017composing}, so they resolve the fraud detection problem, nor the ad matching problem with more media publishers. Most of them also do not offer native support for implementing arbitrary preconditions and postconditions, so agents would need to rely on additional technologies or mechanisms (e.g. legal) to enforce these conditions.

While there also exists secure computation frameworks that are meant to be general-purpose such as ABY, Moose, and MPyC \cite{demmler2015aby, moose, schoenmakers2018mpyc}, users have to write every functionality needed by their sharing problem from scratch, which is cost-intensive. Table \ref{tab:loc} shows the lines of code needed to implement certain computation using a general-purpose secure computation framework. While these functionalities take at most a few lines to implement in Python with an existing Python library or with \textsc{programmable dataflows}, they are complex to implement with secure computation frameworks (e.g. >200 lines to implement set intersection, ~800 lines to implement AES with ABY).

\mypar{Support data preparation} \textsf{upload\_data\_with\_CPR} in line 4-10 from  \F\ref{fig:code_one} checks that all agents' data contains the "CPR" column, before they are uploaded. This specialized \textsf{@api\_endpoint} ensures that all agents' data contains a standard attribute which is used downstream for joining the data. In \F\ref{fig:code_three}, the banks implement a suite of functionalities for data preparation, including \textsf{show\_schema}, \textsf{share\_sample}, and \textsf{check\_column\_compatibility}. With these functions, the banks disclose information to prepare their data in a controlled way, without entering constraint states. 

Support for data preparation is lacking from existing data sharing technologies. For example, a large number of works in Table \ref{tab:technologies} assume that data are horizontal partitions of the same table that share the same schema \cite{bater2017smcql, bater2018shrinkwrap, bater2020saqe, volgushev2019conclave, zheng2021cerebro}. However, in most cases, it is rare that data from different agents are schema-aligned. Hence, these technologies assume the existence of an additional data preparation tools that still need to share data, so not integrating them into the end-to-end data sharing problem is a limitation. 

\mypar{Ensure computational efficiency} Line 13-17 in  \F\ref{fig:code_two} shows how to reuse the combined data from Facebook and YouTube. \textsf{train\_advertising\_model} first checks if the joined data exists. If not, it creates and saves the joined data. If yes, it directly goes to model training. This permits reusing computation.

Such capability is missing from the technologies in Table \ref{tab:technologies}. Most of them use secure multiparty-computation (MPC) for preventing unintended dataflows. To use an MPC technology, agents define their desired computation to run a priori, and the technology constructs a circuit for that computation, such that all agents gets access to the output of the circuit only. Since the entire purpose for these technologies is that agents do not learn anything about each other's data elements other than the output, these technologies offer no support to freely save and reuse parts of the computation. On the other hand, while there exists technologies for running SQL and ML, individually, they cannot be chained together, because no agent can be in possession of the joined results, as that would violate the agents constraints.

\mypar{Effectively manage contract lifecycles} Because agents do not know each other's goals and constraints completely in ad matching and fraud detection, they need to propose and approve contracts on-the-fly. To allow this, they include \textsf{propose\_contract} and \textsf{approve\_contract} explicitly, using the default implementation provided by the CPM, shown in \F\ref{fig:code_two}. On the other hand, DNPR wants to approve all incoming causal queries automatically, so they include the \textsf{upload\_cmr} function to achieve that.

We are not aware of any other technology that enables the kind of interaction facilitated by CPM. Some technologies assume that agents go through an initial phase to agree upon a set of computations to run, without giving sufficient details on how that initial phase can be enabled. For example, Conclave \cite{volgushev2019conclave} assumes that "parties agree via out-of-band mechanisms on the (relational) query to run". Cerebro \cite{zheng2021cerebro} assumes that "parties come together in an agreement phase" to decide on the computation and other parameters of the system, and that "this agreement is enforced by an external mechanism, e.g., through a legal agreement". Many other SQL-based systems \cite{bater2018shrinkwrap, bater2017smcql, bater2020saqe, wang2021secure} assume that the queries to run have already been agreed upon. However, agreeing on the initial phase involves revealing information about agents' data, and it is challenging for agents to reveal sufficient information to each an agreement while avoiding the common constraint states, without support from the sharing technology. Our CPM offers a suite of functionalities for agents to manage the lifecyle of contracts effectively.

\mypar{Summary of Qualitative Results} We show the CPM allows agents to implement complex sharing problems. It supports a wide range of dataflows, permits data preparation, ensures computational efficiency, and manages contract lifecyles effectively.

\subsection{RQ3: Improving Computational Efficiency}
\label{subsec:rq3}

\begin{figure*}[]
  \subfloat{
  \includegraphics[width=0.33\linewidth]{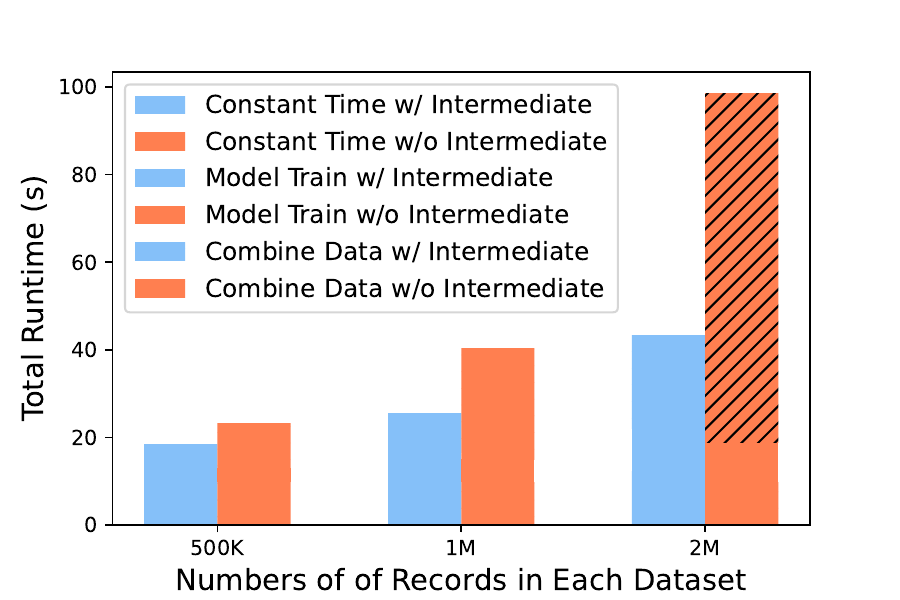}
  \label{fig:intermediate_lr}
  }
  \subfloat{
  \includegraphics[width=0.33\linewidth]{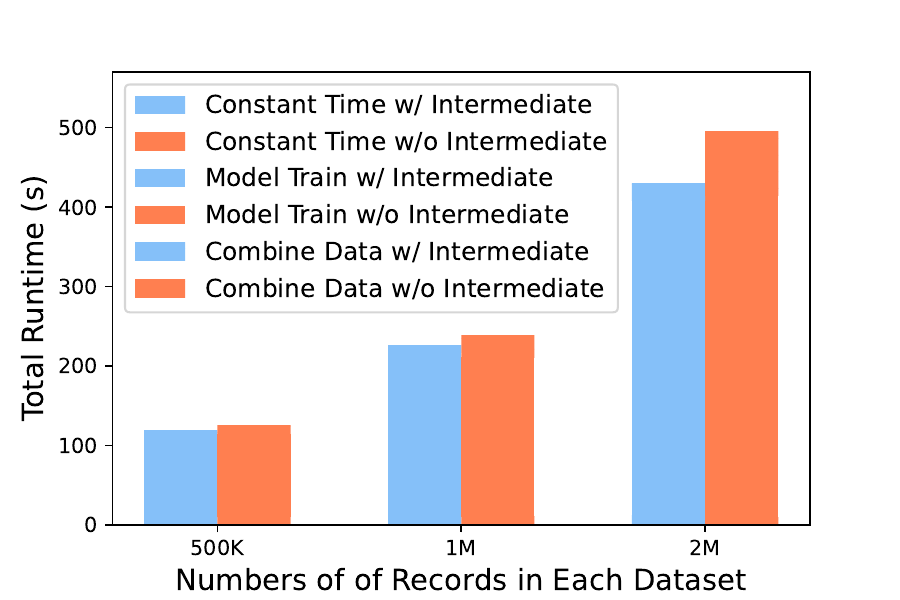}
  \label{fig:intermediate_mlp}
  }
  \subfloat{
  \includegraphics[width=0.33\linewidth]{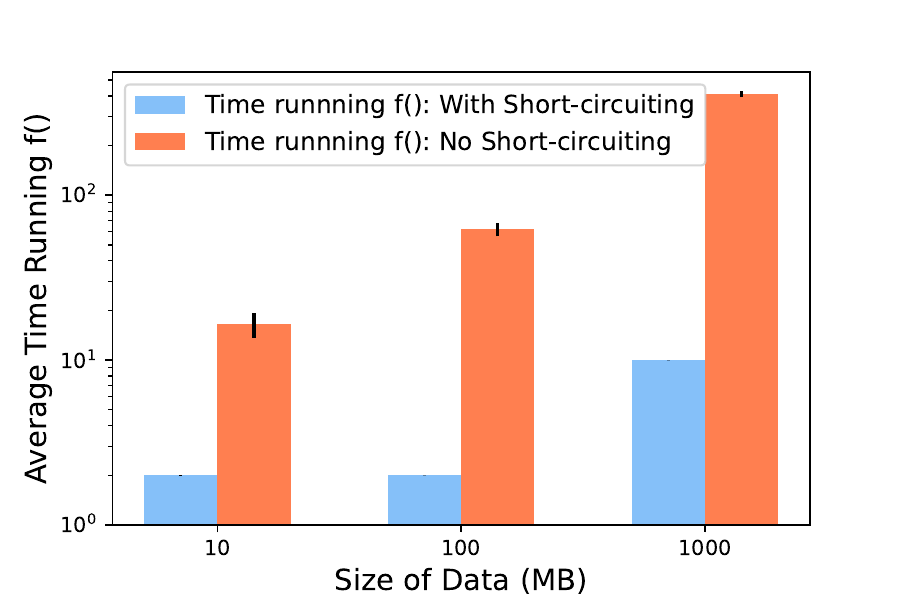}
  \label{fig:shortcircuit}
  }
\caption{Left: Saving Intermediate's Effect in train\_advertising\_model with Logistic Regression. Middle: Saving Intermediate's Effect in train\_advertising\_model with MLP . Right: Short-circuiting's Effect in train\_fraud\_model.}
\label{fig:intermediate}
\end{figure*}

We first study the effect of saving intermediates DEs, and then the impact of short-circuiting on the execution time of contracts. Both extensions lead to significant performance gains.

\subsubsection{Saving Intermediate DEs} As explained in Section \ref{subsec:r3}, the CPM permits reusing intermediate computation results from contracts. Such features are desired in many examples. In the ad matching problem, Don may need to train multiple models over Facebook and YouTube's combined data to obtain a desired one. In a similar problem, there may be multiple analysts who want to train models over the combined data. We study the benefits of reusing intermediates in the ad matching problem.

\mypar{Experimental Setup} We generate Facebook and YouTube's data with a simplified schema based on the actual data Facebook and YouTube collects about its users \cite{facebook_data, google_data}. Both datasets contain attributes including name and email, which will be used as join keys. Each dataset includes some additional attributes, such as age, marriage status, has users liked a Facebook page with games-related content, and has users clicked on some ad displayed on YouTube. 

We compare the performance of two implementations of the the \textsf{train\_advertising\_model}: one uses the programming model's extension to store intermediates, and the other does not. In both implementations, \textsf{train\_advertising\_model} first runs a fuzzy join between Facebook and YouTube's data based on names and email, and then trains either a logistic regression model or a multiplayer perception model over the joined data. Without intermediates, the implementation has to combine the data from scratch every time. With intermediates, the implementation only needs to combine the data the first time it is called, and subsequent calls to this function can skip this step. This implementation is shown in line 13 to 17 in \F\ref{fig:code_two}. We measure the total time it takes for both implementations to run this function multiple times: if the function is only run once, then both implementations would have the same runtime, as they both need to combine the data.

\mypar{Metrics} We evaluate the total time it takes to run the \textsf{train\_fraud\_\\model} function over five runs. The runtime is measured end-to-end.

\mypar{Results} The left and middle subplot of Fig. ~\ref{fig:intermediate} shows the total runtime of the two implementations. The left subplot trains a LR model, and the middle subplot trains a MLP model. The runtime are broken into three parts: 1) \textit{constant time} introduced by the system to run every function, which is fixed at around 2 seconds, 2) time to train the model, and 3) time to combine the data.

The benefits of reusing intermediates for the LR model is clear. When the number of records in each dataset is small (500K), the join finishes relatively fast, so the two implementations have similar runtime. As the number of records increase, the implementation without reusing intermediates take proportionally longer to finish (>2X for 2M records). This large saving comes from the fact that training a LR model is relatively fast comparing to running the join query, so the total runtime is dominated by the time of the join. The more times $f()$ is run, the more significant the saving will become. For the MLP model, there is still runtime saving, albeit less significant than the LR model. This experiment shows that with minimal additional effort from the developers to reuse intermediates (enabling this in ad matching takes 5 lines to implement), agents can implement much more efficient sharing problems. Conversely, an implementation of the contract abstraction without the support for intermediates would lead to repeated computation. If required by some particular sharing problem, agents can even save the final output of a \textsf{@contract\_function}, skipping computation entirely in subsequent calls. 

\subsubsection{Short-circuiting Function Execution} When the data elements accessed by a \textsf{contract\_function} are not part of any approved contract, we want to terminate the execution of the function early to skip computation. We study the efficiency improvement in running the \textsf{train\_fraud\_model} function in the fraud detection example.

\mypar{Experimental Setup} We implement 2 baselines. i) Without short-circuiting. In this case, $f()$ runs to completion, even if it accesses unapproved data elements. ii) With short-circuiting. Here, one bank has an approved contract to access some other banks' data, but the execution of \textsf{train\_fraud\_model} will access data not included in the contract. As soon as unapproved data accesses are detected, the execution stops. The banks train a logistic regression model. We generate the banks' data using a schema from Kaggle. \cite{kaggle}

\mypar{Metrics} We evaluate the average time the system spends on running the \textsf{train\_fraud\_model} function over five runs.

\mypar{Results} The right subplot in Fig. ~\ref{fig:intermediate} shows the average runtime for the baselines for different data sizes. The y-axis is in log scale.

Short-circuiting leads to runtime savings. Over all sizes of data, short-circuiting consistently shows performance gains in orders of magnitudes. Note that, when the data size is small (<= 100 MB), the time spent running $f()$ with short-circuiting is about 2s for every run. This is a parameter set by the system to run the function, before each time data elements access are checked. When the data size is larger (1G), the system spends more time reading the approved data elements, before unapproved data elements are read, so time spent running $f()$ also increases. 

Importantly, in many data sharing problems, the computation agents want to run has a similar structure to the train\_fraud\_model function, where they first read the data from different agents, combine the data, and then perform some computation over the combined data. These computations particularly benefit from the short-circuiting optimization, because access to illegal data elements can be detected before the computation on the combined data element starts, which is typically the bottleneck. On the other hand, if the $f()$ agent processes all the input data elements in a linear fashion, without performing any combining or computation on the combine data, and the illegal data elements are accessed later during $f()$'s execution, the savings will be smaller. 

Overall, the results show that the extensions to our programming model (saving intermediates and short-circuiting function execution) helps agents implement sharing problems efficiently.

\section{Conclusions}
\label{sec:conclusions}

While sharing data increases its value for all agents involved, sharing problems are diverse and have their own bespoke solutions. We create a model that represents any data sharing problem, and derive from it the \textit{contract} abstraction that allows agents to express the intent and understand the consequences of a dataflow, before the dataflow takes place. To permit agents to program their sharing problems with contracts, we implemented a contract programming model. The abstraction and the programming model allow agents to implement complex sharing problems that tailor to their needs.

\bibliographystyle{ACM-Reference-Format}
\bibliography{main}

\end{document}